\documentclass[
    ,final            
  ]
  {aipproc}

\usepackage{graphicx,verbatim,textcomp,amssymb,amsmath}

\layoutstyle{8x11double}

\def\tr{\mbox{tr}}

\begin{document}

\title{Dual condensate and QCD phase transition \footnote{Talk given by Bo Zhang on Quark Confinement and the Hadron Spectrum IX}} 


\author{Bo Zhang }{
  address={Institut f\"ur Theoretische Physik, Universit\"at
  Regensburg, D-93040 Regensburg, Germany}
}

\author{Falk Bruckmann}{
  address={Institut f\"ur Theoretische Physik, Universit\"at
  Regensburg, D-93040 Regensburg, Germany}
}

\author{Christof Gattringer}{
  address={Institut f\"ur Physik, Universit\"at Graz,
Universit\"atsplatz 5, A-8010 Graz, Austria}
} 

\author{Zolt\'{a}n Fodor}{
  address={Department of Physics, University of Wuppertal, Gau\ss{}str. 20, D-42119, Germany}
} 

\author{K\'{a}lm\'{a}n~K.~Szab\'{o}}{
  address={Department of Physics, University of Wuppertal, Gau\ss{}str. 20, D-42119, Germany}
}

\begin{abstract}
The dual condensate is a new QCD phase transition order parameter, which connnects
confinement and chiral symmetry breaking as different mass limits. We discuss
the relation between the fermion spectrum at general boundary conditions and the dual condensate and show numerical results for the latter from unquenched $SU(3)$ lattice configurations.

\end{abstract}

\maketitle


\section{Introduction}

The QCD phase transition manifests itself in two phenomena, deconfinement and chiral symmetry restoration. The conventional order parameter for (de)confinement is the Polyakov loop, the straight loop in the time direction.
After a suitable renormalization, it is related to the free energy of a static quark via $\langle \tr\, P\rangle\sim e^{-\beta F}$ ($\beta=1/k_B T$).
The Polyakov loop is small in the confined phase (and exactly vanishes in the quenched case because of center symmetry) and increases above the critical temperature.

The chiral condensate, 
on the other hand, is an order parameter for chiral symmetry breaking in the massless limit,
as it is not invariant under chiral transformations.
In the chirally broken phase,
the chiral condensate is finite, while it decays above the restoration temperature.

Lattice simulations at physical quark masses have revealed the QCD phase transition 
to be a crossover with pseudo-critical temperatures of $157\pm 4~$MeV for the chiral susceptibility and $170\pm 5~$MeV for the Polyakov loop in \cite{paper:WuppertalPapers}
(see also \cite{paper:hotQCD}).

The dual condensate \cite{paper:QuenchedDualCondensate} 
connects Pol\-ya\-kov loop and chiral condensate as two different mass limits,
thus it also relates confinement and chiral symmetry breaking.
It is therefore particularly interesting to see what one can learn from the dual condensate about the physical crossover.
We here improve previous results \cite{paper:PreliminaryResultUnquenched} and study the dual condensate on the $N_f=2+1$ staggered dynamical configurations of \cite{paper:WuppertalPapers}.

\section{Dual Condensate}

The physical boundary condition for a fermion field at finite temperature is anti-periodic: $\psi(t+\beta,\vec{x})	= -\psi(t,\vec{x})$.
With `quark condensate' we refer to the expectation value $\Sigma(m)=\frac{1}{V} \langle Tr[(m+D_-)^{-1}] \rangle$
with this boundary condition.

We here also consider general boundary conditions \cite{paper:DualCondensate}
\begin{equation}
	\psi(t+\beta,\vec{x})	= e^{i\varphi} \psi(t,\vec{x})\,,	
	\label{eqn:bc}
\end{equation}
giving the general quark condensate
\begin{eqnarray}
	\Sigma(m,\varphi)	= \frac{1}{V} \langle \tr[(m+D_\varphi)^{-1}] \rangle 
	= \frac{1}{V} \sum_{\lambda_\varphi} \frac{1}{m \pm i \lambda_\varphi}\,,	
	\label{eqn:Cond}	
\end{eqnarray}
where $i\lambda_\varphi$ are the eigenvalues of the massless Dirac operator with these boundary condition
(the physical boundary condition $\varphi=\pi$ is among them).

\begin{figure}[h]
\includegraphics[width=0.75\linewidth]{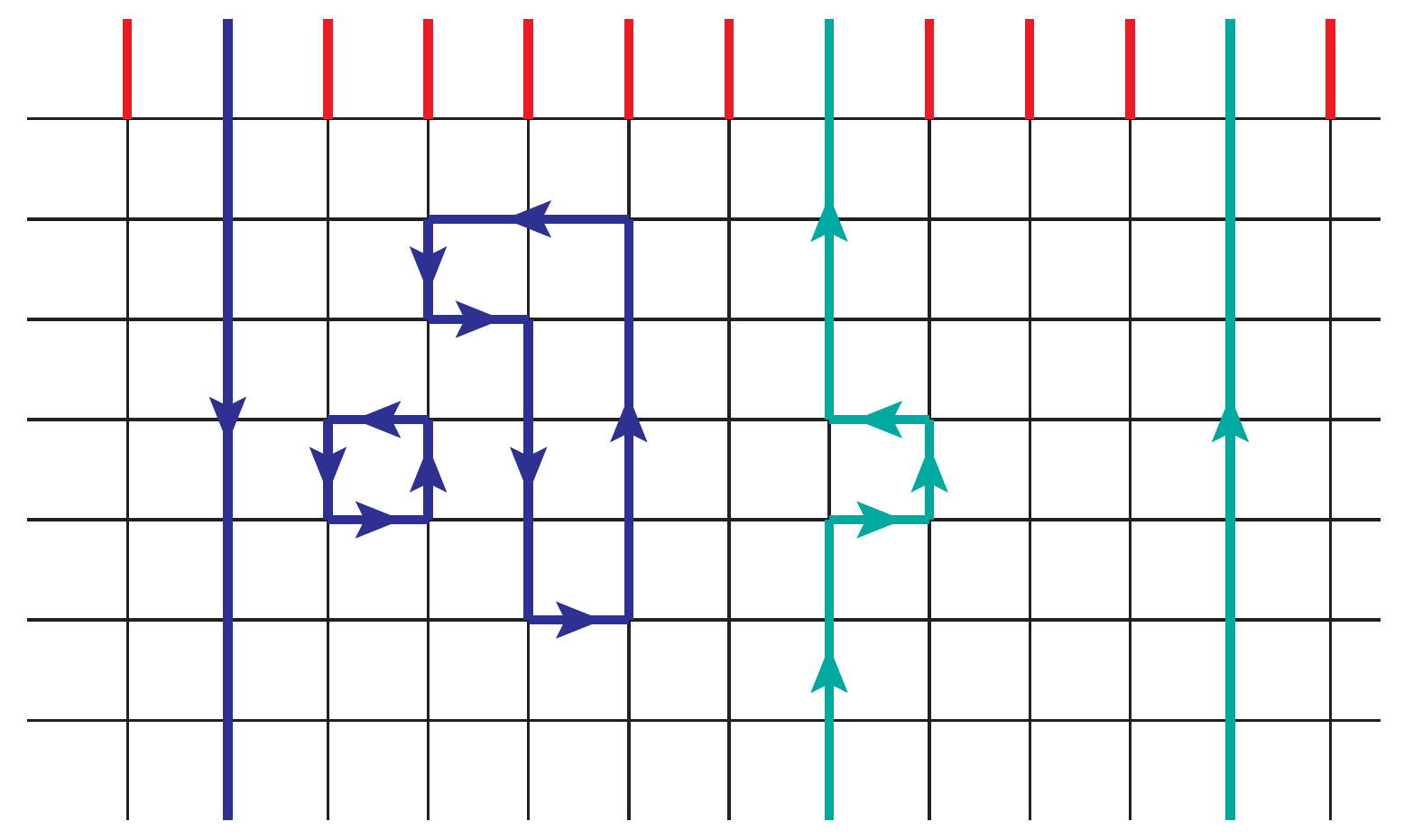}
\caption{Examples of closed loops on the lattice (with time running upwards). The red links get $e^{i\varphi}$ factors
from the implementation of general boundary conditions, Eqn.~(\protect\ref{eqn:implemBCs}). The green lines have winding number one.} 
\label{fig:LoopsInDPhi}
\end{figure}

The dual condensate is defined as the first Fourier component of the general quark condensate 
with respect to the boundary phase $\varphi$:
\begin{eqnarray}
	\tilde{\Sigma}_1(m)=\!	\int_0^{2\pi} \frac{d\varphi}{2\pi}\,  e^{-i\varphi}\,\Sigma(\varphi)
	=\!\int_0^{2\pi} \frac{d\varphi}{2\pi V}
	\sum_{\lambda_\varphi} \frac{e^{-i\varphi}}{m \pm i \lambda_\varphi}\,.
	\label{eqn:DualCond}
\end{eqnarray}

The interpretation of this quantity is simplest in a lattice context. One can implement the boundary conditions (\ref{eqn:bc}) by multiplying a factor $e^{i\varphi}$ to temporal links in one, say the last, time slice
\begin{equation}
	U_0(t=N_t a)  \Longrightarrow  e^{i\varphi}U_0(t=N_t a).
	\label{eqn:implemBCs}
\end{equation}

The general quark condensate $\Sigma(m,\varphi)$ is gauge invariant and as such is composed of 
the contributions from all kinds of closed loops.
These loops receive different powers of $e^{i\varphi}$ factors, see 
Fig.~\ref{fig:LoopsInDPhi}. Now the dual condensate as 
the first Fourier component of $\Sigma(m,\varphi)$ (see (\ref{eqn:DualCond})),
picks out the contributions from all the loops with one $e^{i\varphi}$ factor.
These are loops winding once in the temperal direction, 
hence the dual condensate can be viewed as a `dressed Polyakov loop'.

\begin{figure}[t]
	\begin{minipage}{0.8\linewidth}	
        \hskip0.2cm\includegraphics[width=0.78\linewidth]{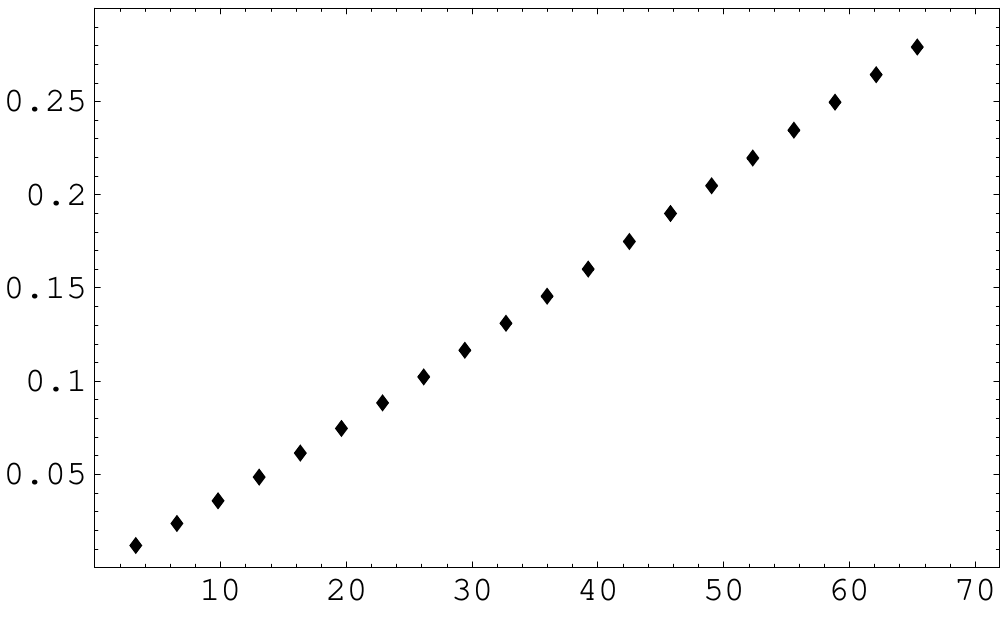}

	\vskip-0.1cm\hspace{2.5cm}$\lambda$[MeV]\vskip0.1cm

\includegraphics[width=0.81\linewidth]{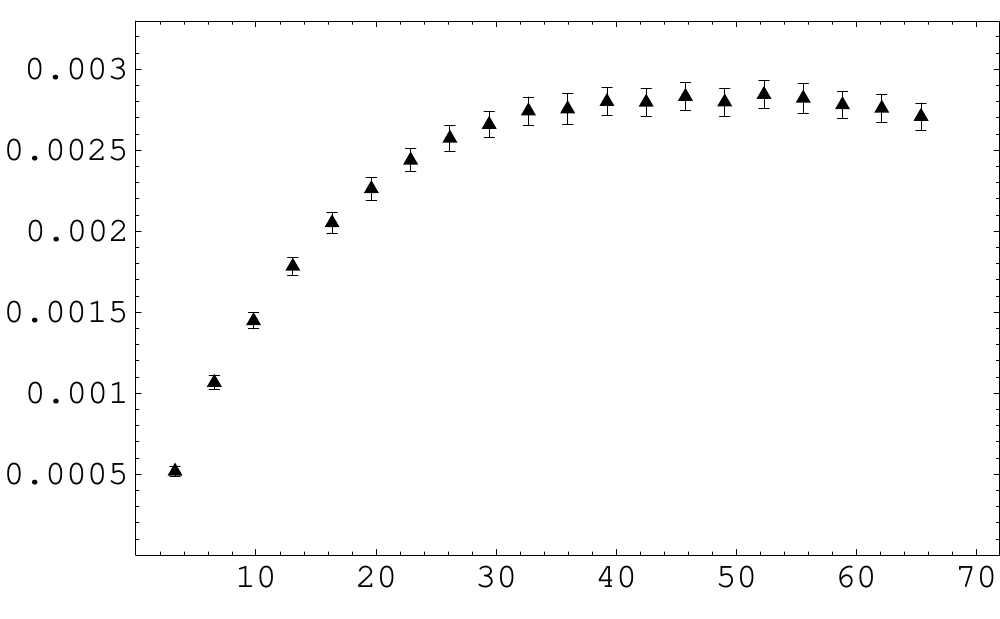}

	\vskip-0.1cm\hspace{2.5cm}$\lambda$[MeV]	

	\end{minipage}
	\caption{Accumulated contributions of eigenvalues to the quark condensate $\Sigma(m,\pi)$ (top) and the dual condensate $\tilde{\Sigma}(m)$ (bottom), both in GeV$^3$, at $T=172~$MeV and $m=100~$MeV.}
	\label{fig:convergence}
\end{figure}

In a similar way dual observables can be constructed for arbitrary gauge invariant objects (cf.~\cite{paper:DualCondensateConvergence}).

The conventional infinitely thin Polyakov loop is included in the set of loops that wind once and dominates in the limit of large probe mass $m$ (which can be seen through an expansion in $1/m$). In this limit, however, more UV eigenvalues contribute to the sum in (\ref{eqn:DualCond}) \cite{paper:QuenchedDualCondensate}. 

We here consider all quantities unrenormalized (in \cite{paper:ProceedingsQuenchedDualCondensate} we demonstrated that the (quenched) dressed Polyakov loop has only a mild dependence on the lattice spacing).

\section{Numerical Results}

We use dynamical improved staggered fermion configurations  
from \cite{paper:WuppertalPapers} 
on lattices of size $8\times24^3$,
for temperatures ranging from $78~$MeV to $890~$MeV
and lattice spacings from $0.282~$fm to $0.028~$fm. 
We compute 500 to 1000 lowest eigenvalues of $D$ for 16 or 8 different boundary conditions
$\varphi\in[0,2\pi]$ with ARPACK.
We currently have completed the spectrum calculations for  20 to 35 configurations at temperatures
between $100~$MeV and $200~$MeV.

\begin{figure}[t]
  \centering
  \begin{minipage}{0.8\linewidth}

	\hskip0.13cm\includegraphics[width=0.782\linewidth]{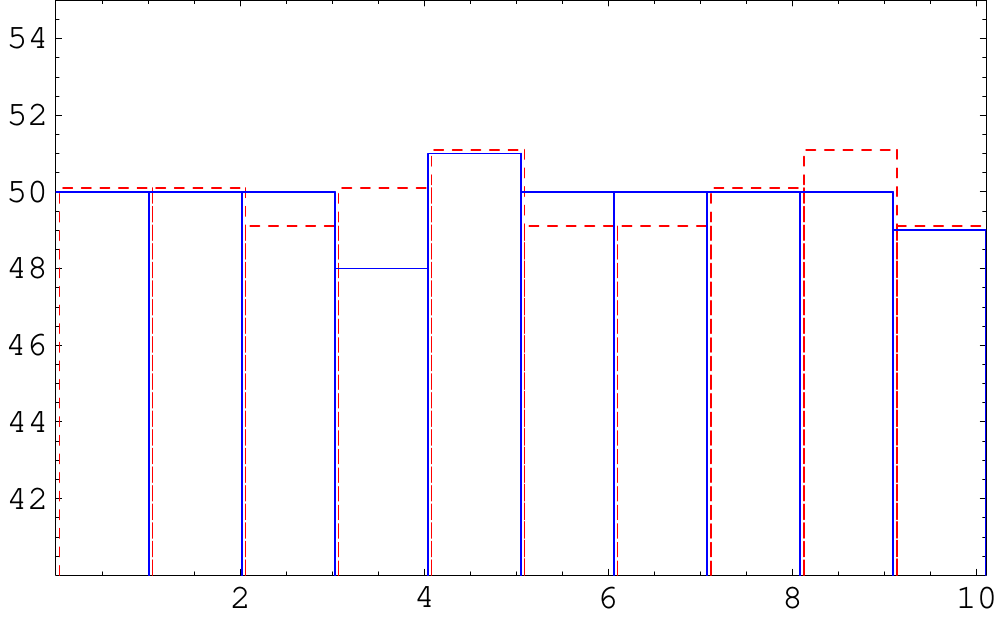} 

	\vskip-0.1cm\hspace{2.5cm}$\lambda$[MeV]\vskip0.1cm

	\includegraphics[width=0.8\linewidth]{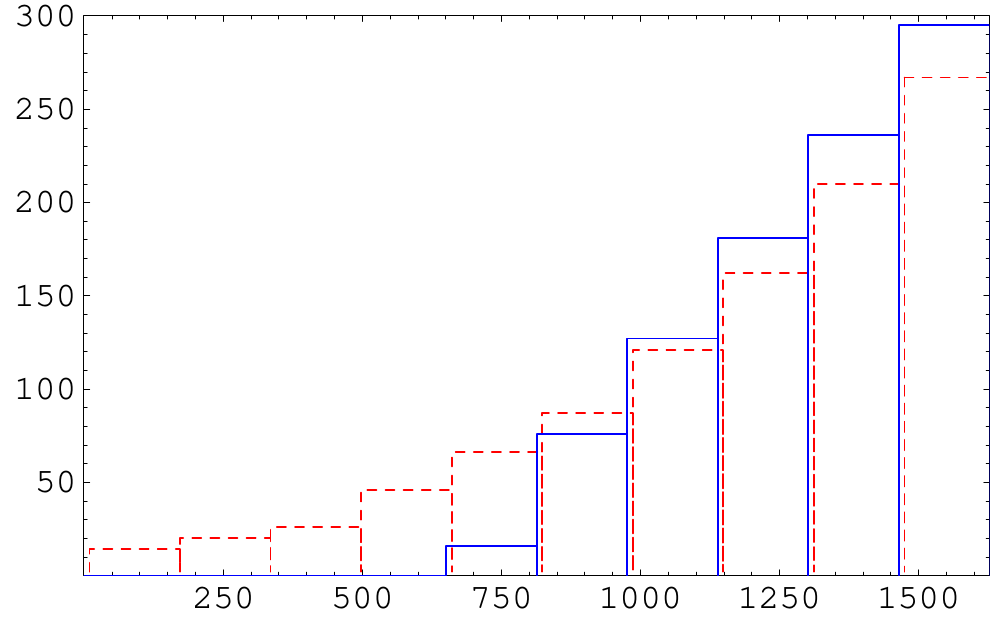}

	\vskip-0.1cm\hspace{2.5cm}$\lambda$[MeV]	

  \end{minipage}
  \caption{Distribution of the lowest eigenvalues for the confined phase ($T=78~$MeV, top) and the deconfined phase ($T=892~$MeV, bottom). We compare histograms for  $\varphi=0$ (red dashed)  and $\varphi=\pi$ (blue).}
  \label{fig:DiracOperatorSpectrum}
\end{figure}

The first problem we investigate is the convergence of the sums (\ref{eqn:Cond}) and (\ref{eqn:DualCond}) when truncated to the number of available eigenvalues (in physical units). 
As the contribution of a $\pm i \lambda$ pair to the condensates 
is $2m/\lambda^2+m^2$, it is clear that
this contribution decays for $\lambda\gg m$ and only the lowest part of the spectrum contributes.

For the dual condensate there is an additional effect because it only probes the difference in the response of the spectra to changing boundary conditions. 
A strong response is manifest only in the IR spectrum
\cite{paper:preQuenchedDualCondensate,paper:DualCondensateConvergence}, 
as can be seen in Fig.~\ref{fig:DiracOperatorSpectrum},
and for dual condensates thus only the IR contributes.
Fig.~\ref{fig:convergence} illustrates this effect: when using the available spectrum, 
the physical chiral condensate from (\ref{eqn:Cond}) has not converged,
while the dual condensate has due to the additional Fourier transform in (\ref{eqn:DualCond}).

\begin{figure}[t]	
	\begin{minipage}{0.95\linewidth}
\centering
	\includegraphics[width=0.8\linewidth]{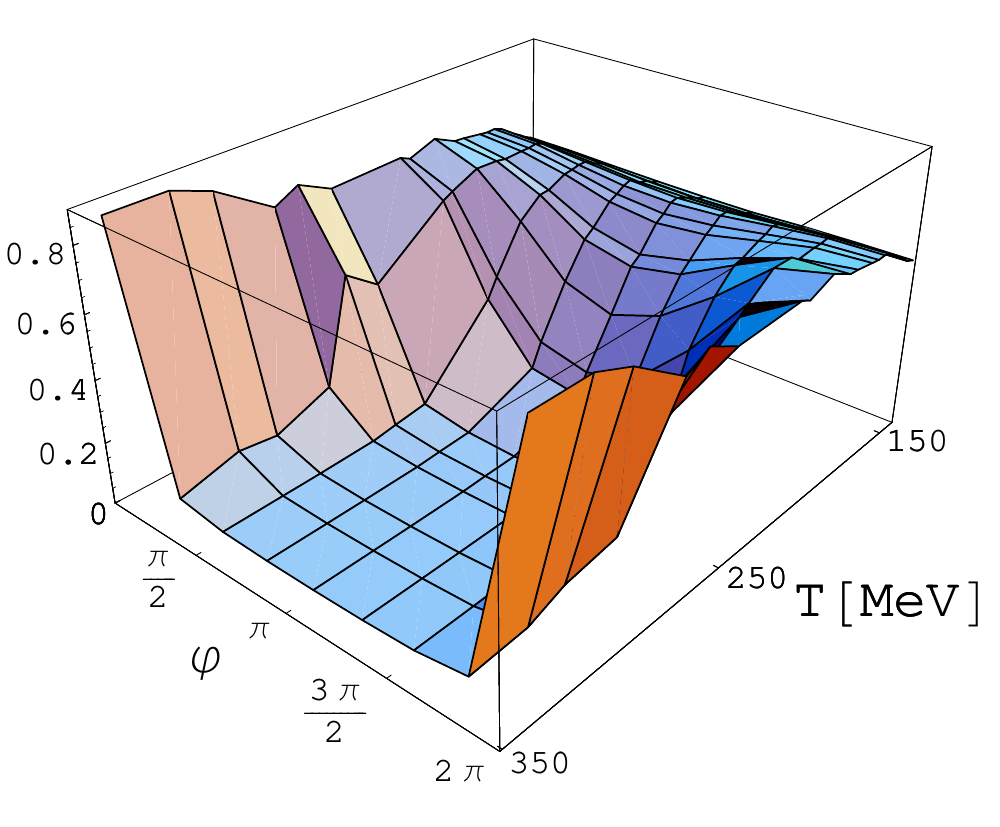}

	\centering
	\includegraphics[width=0.49\linewidth]{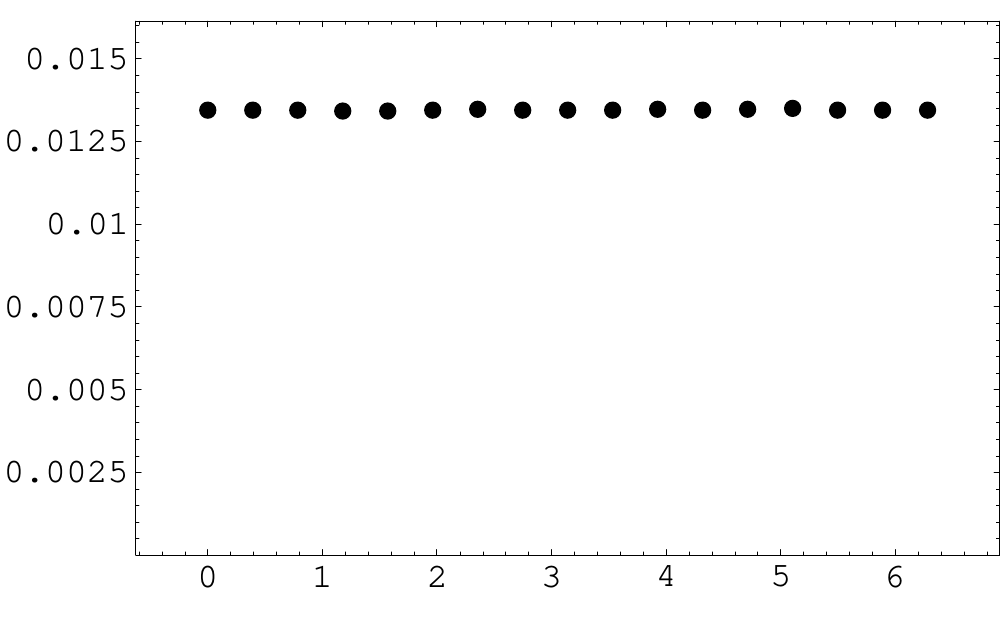}	 
	\includegraphics[width=0.48\linewidth]{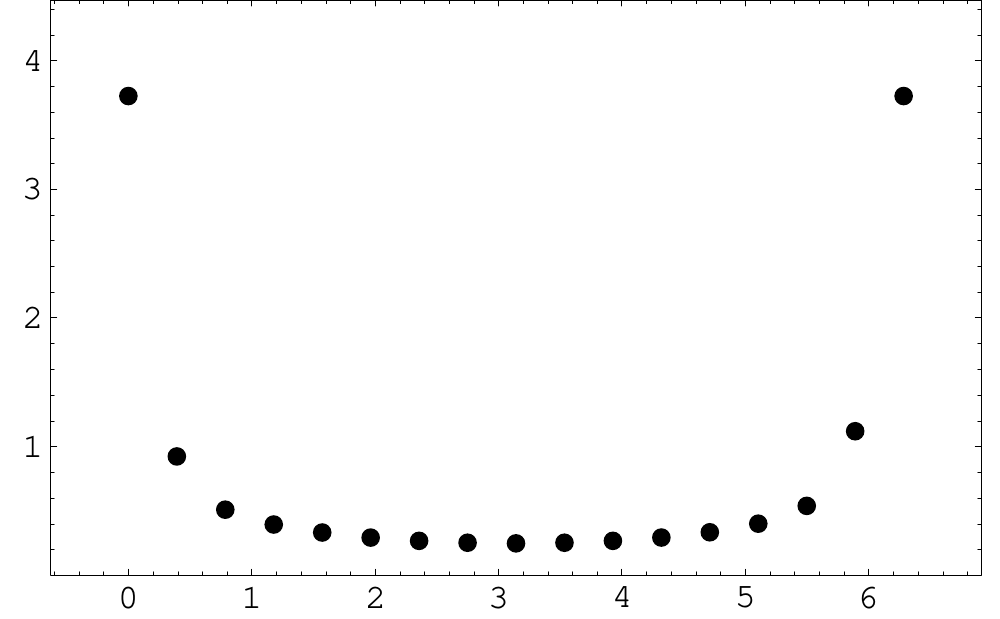}\\ 
	
	\vskip-0.1cm\hspace{0.5cm}$\varphi$	\hspace{3.4cm}$\varphi$
	\end{minipage}

	\caption{The general quark condensate $\Sigma(m,\varphi)$ in $GeV^3$ as a function of temperature and boundary angle with $m=1~$MeV (top). We remark that for large $T$ and $\varphi \sim \pi$ we expect further correlations until full convergence. The lower panels zoom into the confined phase (left, $T=78~$MeV, $m=100~$MeV) and the deconfined phase (right, $T=740~$MeV, $m=10~$MeV), respectively (each for a single configuration).}
	\label{fig:GeneralQuarkCondensateAndBC}
\end{figure}

Fig.~\ref{fig:GeneralQuarkCondensateAndBC} shows the (unrenormalized) general quark condensate as a 
function of the boundary angle $\varphi$ at different temperatures. It is
flat at low temperature and depends strongly on $\varphi$ for high temperatures. Similar results were found also in non-lattice approaches \cite{paper:OtherPapersofSimilarResults}.

\begin{figure}[t]
	\centering
	\begin{minipage}[b]{0.75\linewidth}

		\includegraphics[width=0.9\linewidth]{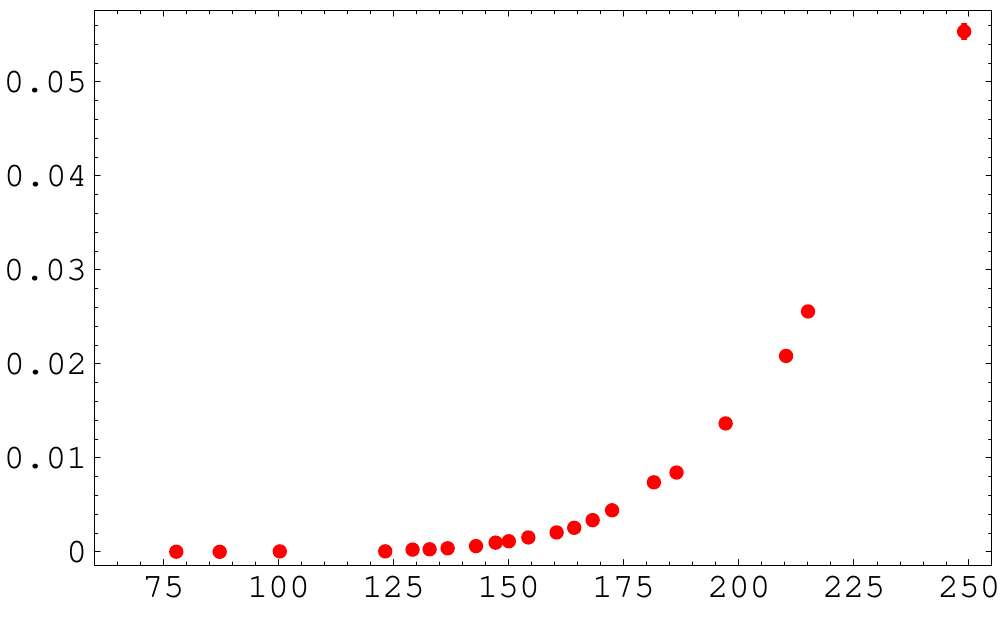}

		\vskip-0.1cm\hspace{2.5cm}$T$[MeV]\vskip0.1cm

		\includegraphics[width=0.9\linewidth]{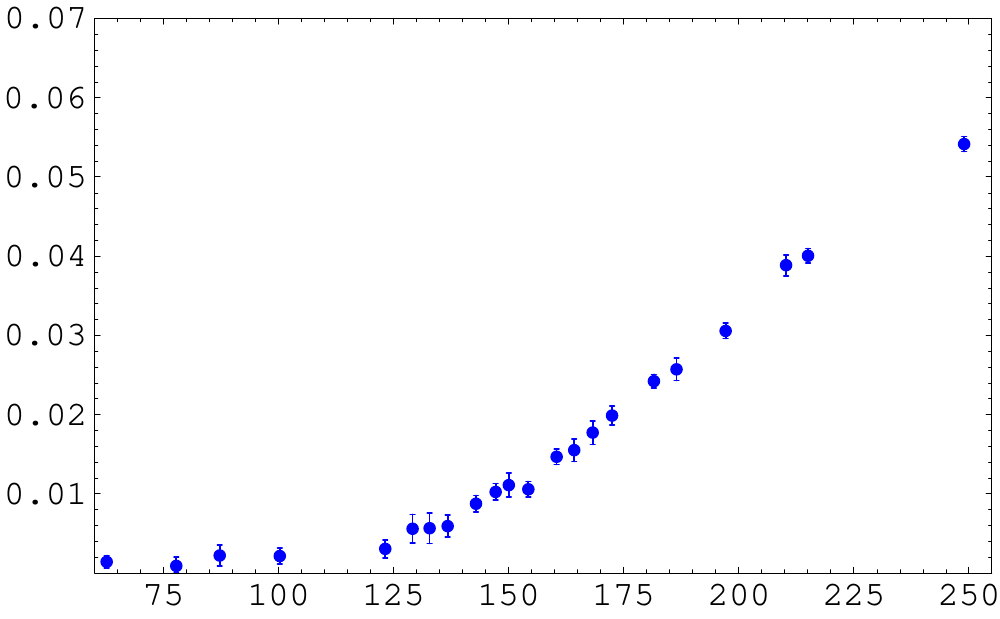}

		\vskip-0.1cm\hspace{2.5cm}$T$[MeV]

	\end{minipage}

	\caption{The dual condensate $\tilde{\Sigma}(m)$ in $GeV^3$ at $m=60~$MeV (top) and the Polyakov loop (bottom) as a function of temperature.}
	\label{fig:DualCondensate}
\end{figure}

\begin{figure}[h]
	\centering
	\begin{minipage}[b]{0.75\linewidth}

		\includegraphics[width=0.92\linewidth]{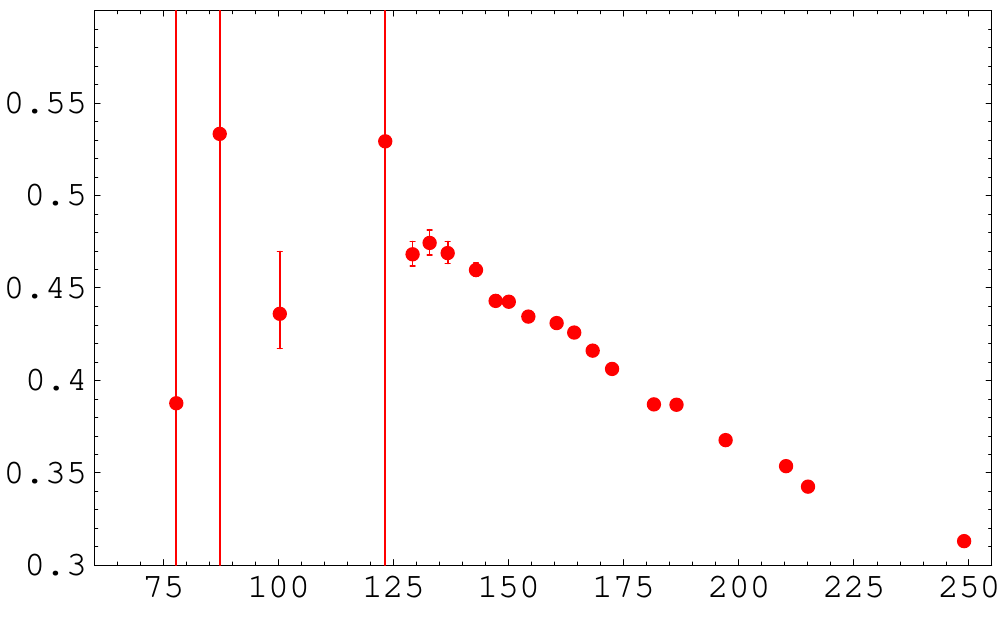}

		\vskip-0.1cm\hspace{2.5cm}$T$[MeV]\vskip0.1cm

		\hskip0.13cm\includegraphics[width=0.9\linewidth]{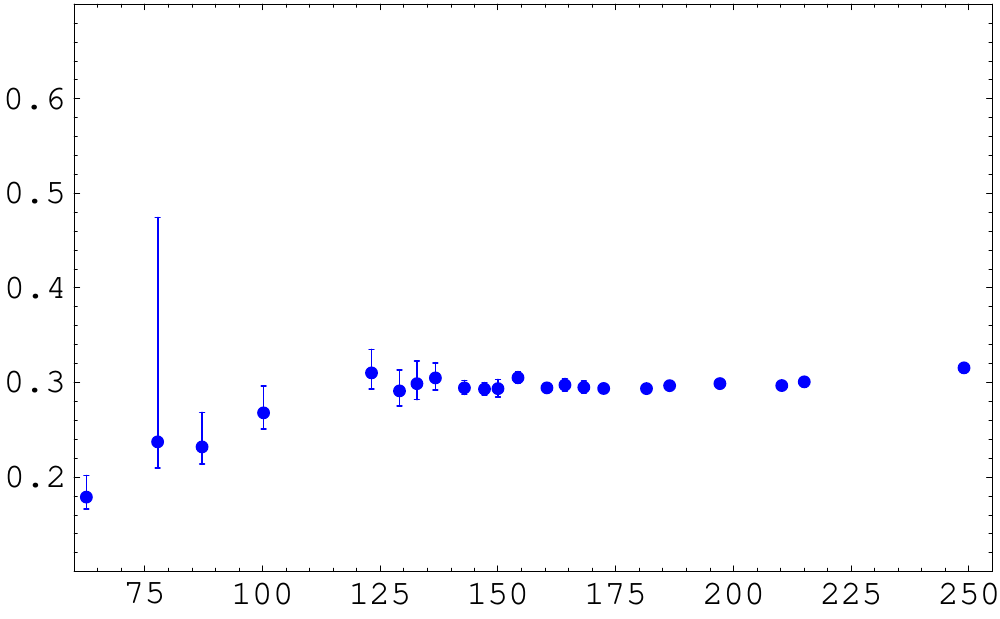}

		\vskip-0.1cm\hspace{2.5cm}$T$[MeV]

	\end{minipage}

	\caption{The `free energy' $-\log \tilde{\Sigma}/\beta$ from the dual condensate at $m=60~$MeV (top) and from the Polyakov loop ($-\log |\langle\tr~P\rangle|/\beta$, bottom) vs.\ temperature.}
	\label{fig:BarePolyakovLoop}
\end{figure}

Correspondingly, the dual condensate is small at low temperature and larger at high temperatures. It should serve as an order parameter for deconfinement, as the Polyakov loop does. In the quenched case this statement can be made exact because of the same behavior under center transformations \cite{paper:QuenchedDualCondensate}. Here both quantities have at least the same qualitative behavior. 

In Fig.~\ref{fig:DualCondensate} we show our results for the absolute value of the unrenormalized dual condensate as a function of temperature, compared to the conventional Polyakov loop. We also plot the negative logarithms of both divided by the inverse temperature (Fig.~\ref{fig:BarePolyakovLoop}). For the Polyakov loop the latter has the interpretation of the free energy of an infinitely heavy quark. In analogy to that we might view the same quantity from the dressed Polyakov loops with mass parameter $m$ as the free energy of a test quark with finite mass $m$.

All of these quantities show an order parameter behavior in the temperature range of $100$ to $200~$MeV. 
In the future we want to identify the critical temperatures (through inflection points and susceptibilities) and study their mass dependence.

We thank Szabolcs Borsanyi for useful correspondence. F.B.~and B.Z.~are supported by DFG (BR 2872/4-2).


\end{document}